\documentclass[aps,prb,twocolumn,showpacs]{revtex4}
\usepackage{txfonts}
\usepackage{graphicx}

\begin{document}

\title{Multi-band effect in the noncentrosymmetric superconductors Mg$_{12-\delta}$Ir$_{19}$B$_{16}$ revealed by Hall effect and magnetoresistance measurements}

\author{Gang Mu, Huan Yang, and Hai-Hu Wen}\email{hhwen@aphy.iphy.ac.cn }

\affiliation{National Laboratory for Superconductivity, Institute of
Physics and Beijing National Laboratory for Condensed Matter
Physics, Chinese Academy of Sciences, P.O. Box 603, Beijing 100190,
People's Republic of China}

\begin{abstract}
We report the longitudinal resistivity and Hall effect measurements
on the noncentrosymmetric superconducting
Mg$_{12-\delta}$Ir$_{19}$B$_{16}$ samples with different critical
transition temperatures. A strong temperature dependence of the Hall
coefficient $R_H$ and nonlinear magnetic field dependence of the
Hall resistivity $\rho_{xy}$ in wide temperature region are
observed, suggesting a strong multi-band effect in this system.
Moreover, a large magnetoresistance up to 20\% is found at the field
of 9 T. We also observe the violation of the Kohler's rule from our
magnetoresistance data, further confirming the presence of
multi-band effect in our samples. A detailed analysis shows that the
data can't be simply described within the two-band scenario at low
temperatures, so we argue that there may be more than two bands
contributing to the conduction of the samples.

\end{abstract}
 \maketitle

\section{Introduction}

The study on superconductors without inversion symmetry has
attracted growing interest in the past
years.\cite{GorkovPRL2001,FrigeriPRL2002,EdelsteinJETP1989,LevitovJETP1985,SamokhinPRB2004,SamokhinPRB2005,Shibayama,Zuev,Samokhin2010}
The penetration depth and NMR
measurements\cite{YuanHQPRL,ZhengGQPRL} in two noncentrosymmetric
superconductors, Li$_2$Pt$_3$B and Li$_2$Pd$_3$B, with similar
structure have shown that the absence of inversion symmetry along
with stronger spin-orbit coupling may allow an admixture of
spin-singlet and spin-triplet pairings. Mg$_{10}$Ir$_{19}$B$_{16}$
(hereafter abbreviated as MgIrB) is another typical material without
inversion symmetry.\cite{Klimczuk} The specific heat and NMR
measurements have shown that the Cooper pairs in MgIrB are
predominantly in the spin-singlet state with an isotropic s-wave
gap.\cite{GangMuPRB2007,MgIrB-NMR} Meanwhile, the unconventional
pairing state is also suggested by the tunneling spectroscopy and
penetration depth measurements in this
system.\cite{Klimczuk2007,Bonalde} The electronic structure of MgIrB
has been calculated using the Korringa-Kohn-Rostoker method showing
a rather complex band structure near the Fermi
surface.\cite{Wiendlocha} Therefore, one may expect a notable
multi-band effect in this system. However, no report about this
issue can be seen from the experimental side. So an in-depth study
using the Hall effect and magnetoresistance (MR) measurements is
required.

We have known that, in conventional metals, the Hall coefficient
$R_H$ is almost independent of temperature in a rather wide
temperature region. However, a clear temperature dependence of $R_H$
have been found in the underdoped cuprate superconductors and
multi-band superconductor
MgB$_2$.\cite{ChienPRB1991,GreenePRL2007,KangAPL2001,KangPRB2002}
The nonlinear magnetic field dependence of the Hall resistivity
$\rho_{xy}$ and rather large MR are also observed in MgB$_2$ thin
films,\cite{HYPRL,QiLiPRL2006} which are attributed to the strong
multi-band effect. Recently, the multi-band behaviors are also
revealed in the newly discovered iron-based superconductors by the
Hall effect measurements.\cite{HQLPhysicaC,PrSr}

In this paper, we report a detailed investigation on the transport
properties on high-quality noncentrosymmetric MgIrB samples with
different critical transition temperatures. We found a strong
temperature dependence of the Hall coefficient $R_H$ and nonlinear
magnetic field dependence of the Hall resistivity $\rho_{xy}$ in
wide temperature region, which were attributed to the multi-band
effect in the system. Moreover, a large magnetoresistance ($\Delta
\rho/\rho_0$) and the violation of the Kohler's rule were observed
in all the samples, further confirming the argument about multi-band
effect.

\section{EXPERIMENT AND SAMPLE CHARACTERIZATION}

The MgIrB polycrystalline samples used in this study were prepared
using a standard method of solid state reaction. The synthesizing
process is the same as our previous work\cite{GangMuPRB2007} except
that a relatively long time up to 10 hours was used in the second
sintering process. The samples were cut into a bar shape with
typical dimensions of 3.2 $\times$ 1.8 $\times$ 0.3 mm$^{3}$. A
six-probe technique was employed to carry out the measurements,
which ensures us to measure the longitudinal and Hall resistivity
simultaneously at each temperature. The magnetic field up to 9 T was
applied perpendicular to the sample surface.

The x-ray diffraction (XRD) measurements of our samples were carried
out by a $Mac-Science$ MXP18A-HF diffractometer with the
Cu-$K_\alpha$ radiation. The ac susceptibility was measured based on
an Oxford cryogenic system (Maglab-Exa-12). The longitudinal
resistivity ($\rho_{xx}$) and the Hall resistivity ($\rho_{xy}$)
were measured on the Quantum Design instrument physical property
measurement system (PPMS) with temperature down to 1.8 K. We used a
dc technique for electrical resistivity measurements and the current
was reversed to correct any thermopower resulting from the
electrical contact. The temperature stabilization was better than
0.1\% and the resolution of the voltmeter was better than 10 nV.

Temperature dependence of the ac susceptibility of three samples
with different $T_c$ are shown in Fig. 1(a). The onset points of the
superconducting transitions in the susceptibility curves are 2.5 K,
4.7 K, and 5.7 K, respectively, for the three samples. The
differences in $T_c$ have been reported to originate from the small
variety of the amounts of Mg and B in the
formula.\cite{MgIrB-NMR,LiZheng} Hereafter we denote the three
samples as MgIrB2.5K, MgIrB4.7K, and MgIrB5.7K, respectively. In
Fig. 1(b), we show the temperature dependence of the longitudinal
resistivity $\rho_{xx}$ near the superconducting transition under
zero field. One can see that the transition width determined from
resistive measurements ($1\% - 99\%\rho_n$ ) is only about 0.2 K,
which is consistent with the rather sharp magnetic transition as
revealed by the ac susceptibility data. The residual resistivity of
about 100 $\mu \Omega$ cm here is much smaller than the value
reported in the previous
work\cite{Klimczuk,GangMuPRB2007,Klimczuk2007}, which suggests that
the samples used in this study is much cleaner with fewer scattering
centers. The resistivity data for two samples MgIrB2.5K and
MgIrB5.7K are shown in the inset in a rather wide temperature range.

The XRD patterns for the two samples MgIrB2.5K and MgIrB5.7K are shown in Fig. 2. It is clear that all the main
peaks can be indexed to the bcc crystal structure. Only tiny peaks from impurities were found, as represented by the blue asterisks.

\begin{figure}
\includegraphics[scale=0.8]{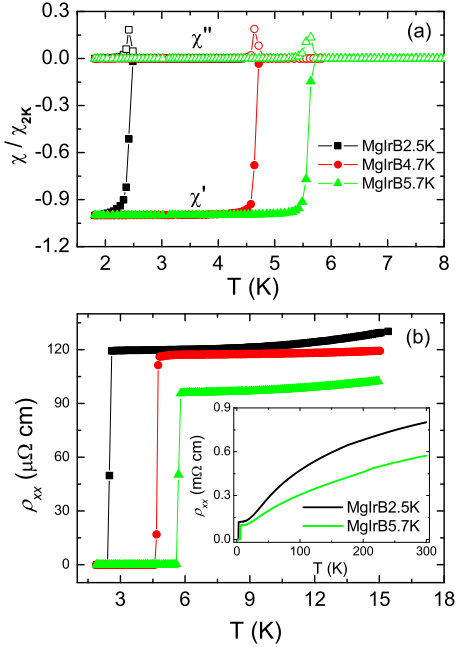}
\caption {(color online) (a) Temperature dependence of the ac
susceptibility of three samples with different $T_c$ measured with
$H_{ac}$=0.1 Oe, $f$ =333 Hz. The curves were normalized by the
value obtained at 2 K. (b) Temperature dependence of the longitudinal
resistivity for the same samples near the superconducting transition under zero field.
Inset shows the resistivity data for two samples MgIrB2.5K and MgIrB5.7K in a wide temperature range up to 300 K.
}
\label{fig1}
\end{figure}

\begin{figure}
\includegraphics[width=0.5\textwidth,bb=10 9 290 240]{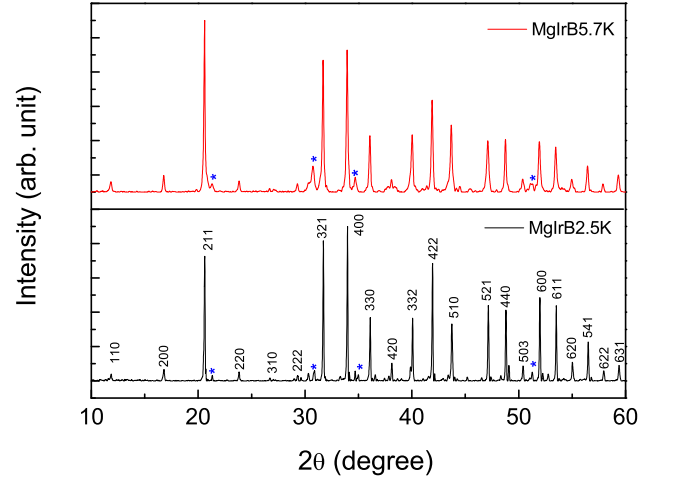}
\caption {(color online) The x-ray diffraction patterns measured for the two samples MgIrB2.5K and MgIrB5.7K.
The peaks from the secondary impurity phase are marked by the blue asterisks. It is clear that the main diffraction
peaks are from the phase of MgIrB.
}
\label{fig2}
\end{figure}

\section{RESULTS AND DISCUSSIONS}

The inset of Fig. 3 shows the field dependence of the Hall
resistivity $\rho_{xy}$ at different temperatures for the sample
MgIrB2.5K. In the experiment $\rho_{xy}$ was taken as $\rho_{xy} =
[\rho(+H)-\rho(-H)]/2$ at each point to eliminate the effect of the
misaligned Hall electrodes. Clear nonlinear field dependence of
$\rho_{xy}$ can be seen in the low temperature region, which is
actually consistent with the multi-band scenario as revealed in
MgB$_2$ and iron-based superconductors.\cite{HYPRL,HQLPhysicaC} As
the temperature increases, the behavior of the curves evolves into
linearity gradually, indicating the weakening of the multi-band
effect at high temperatures. Temperature dependence of the Hall
coefficient ($R_H=\rho_{xy}/H$) for three samples are shown in the
main frame of Fig. 3. One can see rather strong temperature dependence of
$R_H$ in low temperature region. This behavior gives another evidence of the presence
of multi-band effect in our samples. Here we employ a simple
two-band model to interpret our data qualitatively. We have known that
for a two-band system in the low-field limit, the Hall coefficient $R_H$ can be written as
\begin{equation}
R_H=\frac{\sigma_1^2 R_1+\sigma_2^2
R_2}{(\sigma_1+\sigma_2)^2},\label{eq:2}
\end{equation}
where $\sigma_i$ (i = 1, 2) is the conductance of the charge
carriers in different bands, and $R_i = -1/n_ie$ represents the Hall
coefficient for the carriers in each band separately with $n_i$ the
concentration of the charge carriers in the different bands. From
equation (1), we know that the contributions of $R_H$ from different
bands are mainly determined by $\sigma_i$, which can be influenced
by the scattering relaxation time $\tau$. The charge carrier in
different bands may have different $\tau$, which varies differently
with temperature. In this case, each band can produce complex
contributions to the $R_H$. We note that this effect is more
remarkable in the samples with lower $T_c$ and the sign-reversing
effect of $R_H$ can be even seen in the samples MgIrB2.5K and
MgIrB4.7K. This sign-reversing behavior actually indicates the
presence of different types of charge carriers (electron and hole
type) in the present system. The temperature dependence of $R_H$
becomes weak at high temperatures, being consistent with the linear
behavior in the $\rho_{xy}\sim H$ curves at high temperatures as
shown in the inset of Fig. 3.

\begin{figure}
\includegraphics[scale=0.8]{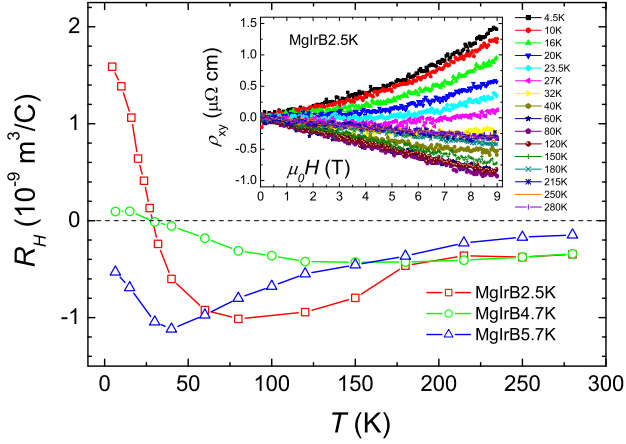}
\caption{(color online) Main frame: Temperature dependence of the
Hall coefficient $R_H$ for three samples MgIrB2.5K, MgIrB4.7K, and
MgIrB5.7K. A strong temperature dependence can be seen below 160 K.
Inset: Magnetic field dependence of the Hall resistivity $\rho_{xy}$ at different temperatures for
MgIrB2.5K.}
\label{fig3}
\end{figure}

To further confirm our conclusions, we study the magnetoresistance
in the same system. Here we present the results of one typical
sample MgIrB2.5K. In Fig. 4(a) we show the temperature dependence of
the longitudinal resistivity $\rho_{xx}$ under different fields. Clear
MR effect can be observed even in this raw data. Here MR is defined as
$\Delta\rho/\rho_0 = [\rho_{xx}(H)-\rho_0]/\rho_0$, where $\rho_{xx}(H)$ and $\rho_0$
represent the longitudinal resistivity at a magnetic field H and that at
zero field, respectively. Temperature dependence of $\Delta\rho/\rho_0$ for two
samples MgIrB2.5K and MgIrB5.7K obtained under 9 T is shown in Fig. 4(b). The value
of MR for the two samples are determined to be 25\% and 20\%, respectively, at about 6 K.
It has been pointed out that in a single-band system, the Lorentz force is balanced by the Hall field.
As a result, the charge carriers can move as if in zero field and they will never be deflected. In that case,
there will be no obvious magnetoresistance observed. So the large value of MR observed here is quite consistent
with the multi-band scenario stated above.

\begin{figure}
\includegraphics[width=0.5\textwidth,bb=6 4 320 255]{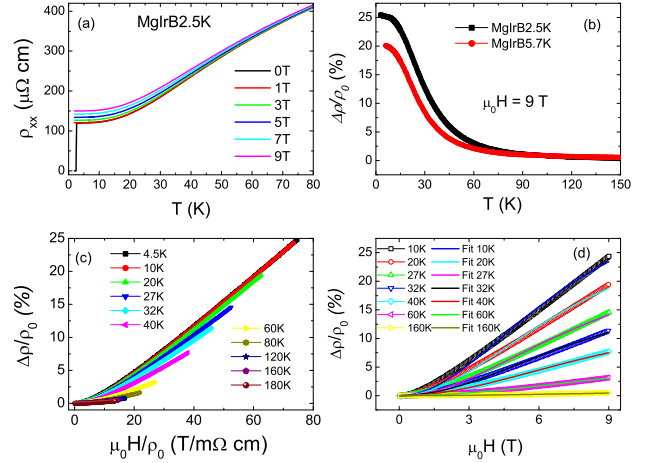}
\caption {(color online) (a) Temperature dependence of the longitudinal resistivity $\rho_{xx}$ for the sample MgIrB2.5K under
different magnetic fields. (b) Temperature dependence of the magnetoresistance $\Delta\rho/\rho_0$ obtained under 9 T for the
samples MgIrB2.5K and MgIrB5.7K. (c) The Kohler plot at different temperatures (see text) for the sample MgIrB2.5K.
(d) Magnetic field dependence of $\Delta\rho/\rho_0$ at several selected
temperatures for the sample MgIrB2.5K. The solid lines represent the theoretical fit to the data using a two-band model.}
\label{fig4}
\end{figure}

Another evidence of the multiband effect is the violation of the
Kohler's rule. The semiclassical transport theory has predicted that
the Kohler's rule will hold if only one isotropic relaxation time is
present in a single-band system.\cite{Kohler} The Kohler's rule can
be written as
\begin{equation}
\frac{\Delta\rho}{\rho_0} =f(\frac{\mu_0H}{\rho_0}),\label{eq:2}
\end{equation}
where $f(x)$ is an implicit function which is independent of
temperature. Equation (2) means that the $\Delta\rho/\rho_0$ vs
$\mu_0H/\rho_0$ curves (the so-called Kohler plot) for different
temperatures should be scaled to a universal curve if the Kohler's
rule is obeyed. Figure 4(c) shows the Kohler plot from the MR data
for the sample MgIrB2.5K. An obvious violation of the Kohler's rule
can be seen from this plot.  We attribute this behavior to the
multi-band effect in this system, just as the case in MgB$_2$, where
the different temperature dependence of the relaxation times in
different bands was considered to be the main reason of the
violation of the Kohler's rule.\cite{QiLiPRL2006}

To get a comprehensive understanding, we have attempted to fit the field dependent MR data using a simple
two-band model:
\begin{equation}
\frac{\Delta\rho}{\rho_0}=\frac{(\mu_0 H)^2}{\alpha+\beta\times
(\mu_0 H)^2},\label{eq:3}
\end{equation}
with $\alpha$ and $\beta$ the fitting parameters which were related
to the conductances and mobilities for the charge carriers in two
bands. Shown in Fig. 4(d) is the magnetic field dependence of
$\Delta\rho/\rho_0$ at several selected temperatures for the sample
MgIrB2.5K. One can see the nonlinear behavior in each curve. The
fitting results using equation (3) are represented by the solid
lines. Clear departure from the experimental data can be observed in
the low temperature region. When the temperature is higher than 60
K, the theoretical curves based on the two-band model can roughly
describe the data. Therefore, we can conclude that, at least in low
temperatures, there may be more than two bands contributing to the
conduction of our samples. In other words, the conduction of our
samples is dominated by more than two bands at low temperatures,
then it behaves like the two-band scenario in the intermediate
temperature region, and finally it evolves to the single-band
behavior at temperatures above 160 K when the interband scattering
becomes very strong.

\section{CONCLUSION}
In summary, Hall effect and magnetoresistance were measured on the
noncentrosymmetric superconductors Mg$_{12-\delta}$Ir$_{19}$B$_{16}$
with different $T_c$. We found a strong temperature dependence of
the Hall coefficient $R_H$ and nonlinear field dependence of the
Hall resistivity $\rho_{xy}$ in wide temperature region in all the
samples investigated here. Moreover, a rather large
magnetoresistance up to 20\% under 9 T and the violation of the
Kohler's rule were observed in our samples. These experimental
features are all consistent with the conclusion that the samples we
studied belong to the multi-band system. A more detailed analysis
shows that there may be more than two bands contributing to the
conduction of the samples in the low temperature region. It is this
multi-band effect that weakens the expected admixture of the
spin-singlet and spin-triplet pairings.

\begin{acknowledgments}
This work is supported by the National Science Foundation of China,
the Ministry of Science and Technology of China (973 project,
Contracts Nos. 2006CB601000 and 2006CB921802), and Chinese Academy
of Sciences (project ITSNEM).
\end{acknowledgments}

\end{document}